\documentclass[12pt,reqno]{amsart}
\usepackage{graphicx}
\usepackage{caption}
\usepackage{textcomp}
\usepackage{setspace}
\begin{document}
\title [CES v. LS]{Correlation Estimation System Minimization Compared to Least Squares Minimization in Simple Linear Regression }
\author{Rudy A. Gideon}

\begin{abstract}
\noindent A general method of minimization using correlation coefficients and order statistics is evaluated relative to least squares procedures in the estimation of parameters for normal data in simple linear regression.
\end{abstract}
\maketitle
\section{Introduction}

In Gideon and Rothan (2011), the use of correlation coefficients (CC) was related to existing optimal linear estimation of variation or scale on ordered data. Another paper, Gideon (2012), gives Correlation Estimation System (CES) examples in many areas: simple linear regression, scale equations, multiple linear regression, CES minimization to estimate $\sigma_{res}/\sigma_y$, nonlinear regression, and estimation of parameters of univariate distributions. CES is a general system of estimation using CCs as the starting point. In Gideon and Rothan (2011), Pearson's CC was related to linear estimates of variation based on order statistics. This paper uses Pearson's CC on order statistics in the simple linear regression model $E(Y|x)=\beta*x$  to see how it compares to classical least squares. By simulation it is found that this estimation technique almost duplicates classical least squares regression results. More importantly, any CC, such as those in Gideon (1987 and 2007), could be used (even those based on ranks) to do simple linear regression (SLR); the CES minimization technique gives a very general way of allowing any correlation coefficient to tackle a wide variety of regression problems.

\section{The Minimum Slope Method}
The CES minimization technique is developed in this section for a random sample $(\underline{x},\underline{y})$. Assume without loss of generality the intercept is zero so as to better visualize the regression of $(\underline{y}-b\underline{x})^0$ on $\underline{x}^0$ for a selected $b$. The slope of this regression is denoted $s$. The superscript $0$ indicates that the elements of the vector are ordered from least to greatest. It is critical to observe that the elements of the residual vector are not paired with specific components of $\underline{x}$. After selecting a value of $b$, plot $(\underline{y}-b\underline{x})^0$ versus  $\underline{x}^0$. If $b$ is such that $b\underline{x}$ produces  residuals with wide variation, the $(\underline{y}-b\underline{x})^0$ versus $\underline{x}^0$ plot is steep and the regression has a large slope $s$. As $b$ approaches a more reasonable value, the residuals, $\underline{y}-b\underline{x}$, become closer in value and the  $(\underline{y}-b\underline{x})^0$ versus $\underline{x}^0$ plot is less steep and so $s$ is smaller, but always nonnegative due to ordering the residuals. If, of course, $\underline{y}=b\underline{x}$ the vector $\underline{0}$ regressed on $\underline{x}^0$ gives $s=0$. By choosing the $b$ that minimizes slope $s$ (the Minimum Slope method or MS) the residuals $\underline{y}-b\underline{x}$ are as uniform as possible, i.e. they increase as little as possible from least to greatest. So MS regression is a sort of minimum totality criterion. The CES-MS  can use any CC to find the $s$ -- each may give a different value -- but with "good" data they are all close to each other.
Recall that in SLR,  $r_p(independent variable,\; residuals)=0$, where  $r_p$ is Pearson's correlation coefficient. Analogously, let the independent variable be $\underline{x}^0$ and let the residuals be $(\underline{y}-b\underline{x})^0-s\underline{x}^0$ to obtain the equation
\begin{equation}  r_p(\underline{x}^0,(\underline{y}-b\underline{x})^0-s\underline{x}^0)=0.
\end{equation}
In this equation $r_p$ is used, but any CC could be employed instead. To obtain the estimation, select the value of $b$ which minimizes $s$.
 This equation is set up in R code in Section 4 and is generally solved by an iterative technique. For Pearson's CC the solution is, for a selected $b$,
 $s=\dfrac{\Sigma (x_{(i)}-\bar{x})res_{(i)}}{\Sigma( x_{(i)}-\bar{x})^2 }$, where $res_{(i)}$ is the $i^{th}$ smallest element of $\underline{y}-b\underline{x}$; $s$ estimates $\sigma_\epsilon/\sigma_x$.
The intercept is chosen by using a location statistic on $(\underline{y}-b\underline{x})$.

In CES, SLR analysis estimates of $\sigma_y$ and $\sigma_x$ are needed. To estimate these scale quantities solve for the slope, $s$,  in the equation
\begin{equation}  r_p(\underline{e},\underline{x}^0-s\underline{e})=0
\end{equation}
where $\underline{e}$ is the vector with components $\Phi^{-1}(i/(n+1)),\; i= 1,2,\dots,n$.  $\Phi$, the distribution function of a standard normal random variable, is used because the simulations are from the normal distribution.
Preferably $\underline{e}$ should be the expected values of the order statistics, but they are not available for all sample sizes, and so are replaced by estimates that converge to them.

Simulations were run using the random number generator of R to generate random samples for the bivariate normal distribution.
The five parameters of the bivariate normal were chosen with the means always zero and then random samples drawn.  The conditional distribution of $Y$ given $x$ was studied.  The slope parameter, $\beta$, and the standard error, $\sigma_\epsilon$, were calculated for the random samples. These CES methods have been investigated in more complex situations, as stated above, all with excellent results: Gideon (2012) and some unpublished research, including a paper on the Logistic Distribution as well as Sheng (2002) on Time Series.

For each random sample the parameters of the bivariate normal distribution and the regression parameters of the conditional distribution of Y given x were estimated by two methods:

 (a) least squares or classical normal theory and

 (b) Correlation Estimation System, CES, using the minimization equation (1).

  The most surprising result in comparing the methods (a) to (b) is that (b) is as good as classical normal theory when the Pearson's correlation coefficient is used.  The main point of this paper is to show the apparent equivalence, in the sense of equal in distribution (Randles and Wolfe, 1979), in simple linear regression between LS (least squares minimization) and CES using Pearson's correlation coefficient. This technique, when used with robust CCs, provides a very simple way to judge the regression without having to evaluate and perhaps delete data points. When the three robust CCs, Greatest Deviation, Absolute Value, or  Median Absolute Deviation, (these are defined in Gideon (2007) and are used in Table 2) are used over time in simple linear regression and compared to the LS line, it quickly becomes apparent what sort of data causes the robust lines to be different from the LS line, even though the data may not have "real" outliers. The example in Table 2 illustrates this. There are also some data sets in which certain points are considered outliers when they are actually not. Using the robust CCs allows these points to have a role in the analysis without overwhelming it.

To compare the two systems, each process is run on the same data.
There are two primary comparisons:

(1) $\hat{\sigma}_\epsilon(LS )$ and $\hat{\sigma}_\epsilon(CES)$, the standard deviations of the LS and MS residuals (the sum of squares of deviations from the mean divided by $n-1$) are computed and compared.

(2) Let $\hat{\sigma}_{ratio}$ represent the MS estimate of $\sigma_\epsilon/\sigma_x$.
Now the ordered residuals from least squares can be regressed against $\underline{x}$, using equation (1)  with the LS estimate of  slope $\beta$ to find $s$. Doing this shows how well LS minimizes the slope so it can be compared to the MS result. Let $\hat{\sigma}_{LSratio}$ represent this estimate of $\sigma_\epsilon/\sigma_x$.

(2a) Let $\hat{\sigma}_x(CES)$ denote the estimate of $\sigma_x$ obtained by using equation (2).  Now multiply this result by  $\hat{\sigma}_{ratio}$, to obtain  estimate $\hat{\sigma}_\epsilon(MS)$, an estimate of $\sigma_\epsilon$. $\hat{\sigma}_{LSratio}$ is multiplied by  $\hat{\sigma}_x(CES)$ to obtain $\hat{\sigma}_\epsilon(LS2)$, another estimate of  $\sigma_\epsilon$.

Some insight for the estimator of $\sigma_{ratio}$ comes from Gideon and Rothan (2011). Let the random variable Z be N(0,1), U be N(0, $\sigma_u^2$), and T be N(0, $\sigma_t^2$). Using order statistics notation let, $u_i=E(U_{(i)})$, $ t_i=E(T_{(i)})$, and $k_i=E(Z_{(i)}),
i= 1,2,\dots,n$. Now let  $\underline{u}$ and  $\underline{t}$ be the vectors of the expected values of these order statistics.
The equation  $r_p(\underline{u},\underline{t}-s\underline{u})=0$ is the same as (2) but with limiting values substituted for the data.

Now $u_i=E(U_{(i)})=\sigma_u*k_i$ and $t_i=E(T_{(i)})=\sigma_t*k_i$ and so the solution to the equation is $s=\dfrac{\Sigma u_it_i}{\Sigma u_i^2 }$ $=\dfrac{\Sigma (\sigma_uk_i)(\sigma_tk_i))}{\Sigma (\sigma_uk_i)^2}=\dfrac{\sigma_t}{\sigma_u} $.
The fact that this estimation concept works on data is illustrated in Table 1.

For the least squares residuals, solving for $s$ in $r_p(\underline{e},\underline{res}^o-s\underline{e})=0$  estimates $\sigma_\epsilon$ because $\sigma_z=1$.
Since $\underline{e}=\Phi^{-1}(\underline{p})$, where \\
$\underline{p}= (1/(n+1), 2/(n+1),\dots, n/(n+1))\prime$, the elements of $\underline{e}$ approach the expected values of the order statistics, that is, $\Phi^{-1}(p_i)$ approximates $E(Z_{(i)})$. This can be seen in Table 1; as the sample size increases, $\hat{\sigma}_x(CES)$ approaches $\sigma_x$.

As already explained, solving for $s$ in $r_p(\underline{x}^o,\underline{res}^o-s\underline{x}^o)=0$ estimates
$\sigma_e/\sigma_x$, i.e. $\hat{\sigma}_{LSratio}$, and the solution to equation (1) for $r_p$ is  \\
$s=\dfrac{\Sigma (x_{(i)}-\bar{x})res_{(i)}}{\Sigma( x_{(i)}-\bar{x})^2 }$. The reasonableness of this process is shown by replacing data with theoretical counterparts. Thus, the $x_{(i)}-\bar{x}$ terms are replaced by $E(X_{(i)}-\bar{X}) =E(X_{(i)})-\mu_x=\sigma_x k_i$. The term $res_{(i)}$ is $Y_j-bx_j$ for some $ j$, $1\le j \le n.$ The conditional distribution of $Y_j|x_j$ is
$N(\frac{\sigma_y}{\sigma_x}\rho x_j,\; \sigma^2_y(1-\rho^2))=N(\beta x_j,\; \sigma^2_\epsilon)$, or
$Y_j|x_j-\beta x_j$ is $N(0,\; \sigma^2_\epsilon)$. Thus, for each $i$ there is a $j$ such that
$res_{(i)}=(Y_j-bx_j)_{(i)}$ and $(Y_j-bx_j)_{(i)}$ is replaced by $E(Y_j|x_j-\beta x_j)_{(i)}=\sigma_\epsilon k_i$. So $s$ now becomes
$\dfrac{\Sigma (\sigma_x k_i)(\sigma_\epsilon k_i))}{\Sigma (\sigma_x k_i)^2}=\dfrac{\sigma_\epsilon}{\sigma_x} $, as expected.

\section{Results}
The Tables are representative examples of the many simulations used to study the MS technique. Table 1 results are averages; an individual sample analysis helps put the table in perspective and also helps give meaning to the notation. Start with one sample of size 25 with the same parameters as the third set in Table 1, namely:
$$  \rho =0.5727, \beta =0.8008, \sigma_\epsilon =1.7659, \sigma_x =1.5403, \sigma_y =2.1541.$$
The two estimated slopes were found to be $\hat{\beta}(LS)=0.6785$  and  $\hat{\beta}(MS)=0.6051$; typically, the estimated slopes are close, but in this example they are somewhat different. The intercept used for the CES method comes from an unpublished paper deriving a location estimator from the Greatest Deviation CC; it is essentially the average of the first and third quantiles of the sample. The intercepts are $int(LS)=-0.3001$ and $int(MS)=-0.3571$. The two sets of residuals are compared using the standard deviation formula: $\hat{\sigma}_\epsilon(LS)=1.4736$ and  $\hat{\sigma}_\epsilon(CES)=1.4796$. Note that the denominator is $n-1$ rather than the usual $n-2$, and that the LS quantity is just barely smaller than the CES-MS value. Now use the LS residuals in the MS method to see how it compares to the MS minimum of $\hat{\sigma}_{ratio}=0.7822$. The LS residuals produce a value of $\hat{\sigma}_{LSratio}=0.7836$. Here LS has the slightly higher value. These results were consistent over many samples; there was very little variation between these two quantities within a sample and almost always   $\hat{\sigma}_\epsilon(LS)$ was barely smaller than $\hat{\sigma}_\epsilon(CES)$. Likewise, $\hat{\sigma}_{ratio}$ was just barely smaller than $\hat{\sigma}_{LSratio}$. Now these last two values are multiplied by  $\hat{\sigma}_x(CES)=1.9440$ to obtain $\hat{\sigma}_{\epsilon}(LS2)=1.5234$ and  $\hat{\sigma}_{\epsilon}(MS)=1.5207$. Both values are very close but somewhat different from the LS estimate of $\sigma_\epsilon$, 1.4736. Finally, $\hat{\sigma}_x(LS )=1.8180$. For small sample sizes the bias in the CES method for $\sigma_x$ makes it larger than the classical estimate. It was also true that the two estimates of
$\sigma_\epsilon$ within each method, both (1) and (2), were always close together and much further apart between methods. Each set of residuals had 13 negative values and each had 12 positive values. In Table 1 it is unclear which method gives averages closest to $\sigma_\epsilon$.

The following observations come from Table 1. As the sample size increases both methods get more accurate without one being better than the other. For all sample sizes and parameter values,  $\hat{\sigma}_\epsilon(LS )$ is always better than $\hat{\sigma}_\epsilon(CES )$, but only by a very small amount. On the other hand,  $\hat{\sigma}_{ratio}$ is generally better than  $\hat{\sigma}_{LSratio}$, again  by a very small amount. The few cases in which $\hat{\sigma}_{ratio}$ is not better have high correlations. However, in these cases both systems are essentially indistinguishable. The general conclusion is that fundamentally MS and LS give essentially the same minima for each sample; however, the values of the residuals can be slightly different.  The average values of the MS and LS minima also show a very small difference, one that usually affects at most the value of the least significant digit of the data.

The whole point so far has been showing how the CES method with $r_p$ using MS is essentially equivalent to classical normal theory. The MS method is now used with other CCs.
First, the R code given in Section 4 uses $rfcn$ as a generic symbol to be substituted everywhere for $Pfcn$ in the code. There are no changes in the R code except to define the CC via $rfcn$.  So far only $rslp = Pfcn$ was used to assign Pearson's CC to be employed in the MS technique. In Table 2, $rfcn$ was assigned to be, in order,  the Greatest Deviation CC, $GDfcn$; Kendall's $\tau$, $Kenfcn$; Gini's CC, $Ginfcn$; the Absolute Value CC, $absfcn$; the Median Absolute Value CC, $madfcn$. The program $Cordef\&reg.R$ on the website has all of these functions. This R program includes a tied value procedure for rank based CCs.

The first row of the Table 2 shows outcomes as in Table 1, that is, LS compared to CES with Pearson's CC. This is a different run than the earlier one sample example, but the results are very similar; the slope estimates are very close, and the two minimizations give comparable results as before. The other five CCs give good estimates of $\beta$. In column two are the results of the LS idea of sum of squares of the residuals. This column contains the square root of the sum of squares divided by 24, \textit{i.e.} $n-1$. If this is changed to the unbiased quantity by multiplying by 24/23,  results closer to $\sigma_{\epsilon} $ are obtained. Finally, the CES-MS results are in column three and in all cases the CES method gives a lower minimum than the LS2 method directly above.

Two of the robust CCs, GDCC and the Absolute Value CC, are much closer to $\beta$ than LS; in column 3 the $\hat{\sigma}_\epsilon(MS)$ values of these two CCs are the smallest. Several $x$-points (near the maximum and minimum $x$-values) have $y$-values that are high or low enough to unduly influence the LS method to increase the slope estimate. For real problems with unknown parameters observational experience on fitting CES lines and comparing to LS results soon leads one to recognize when LS may not be "best." In Table 2 notice that all CES CCs have $\hat{\sigma}_\epsilon(CES )$ closer to $\hat{\sigma}_\epsilon(LS)$ than $\hat{\sigma}_\epsilon(LS2)$ is to $\hat{\sigma}_\epsilon(MS)$.

 \begin{table}
\caption*{Table 1: Comparison of  Two Minimization Processes}
\vspace*{-.3cm}
\centerline{All Tabular Entries are Means}
\vspace*{.5cm}

\begin{center}
\begin{tabular}{|l|l|c|c|c|}
\hline

\multicolumn{5} {|c|} {  $\rho=0.9216$  $\beta=0.3800$ $\sigma_{\epsilon}=0.8000$    } \\
\multicolumn{5} {|c|} {$\sigma_x=5.000$ $\sigma_y=2.0615$ }\\  \hline
  & nsim=100 & n=20& n=50 & n=100 \\ \hline
slopes  & $\hat{\beta}(LS)$& 0.3839& 0.3740& 0.3812 \\
&$\hat{\beta}(MS)$& 0.3833& 0.3742& 0.3808 \\ \hline
minima& $\hat{\sigma}_\epsilon(LS )$ &0.7624&0.7811&0.7893 \\
(LS Method)& $\hat{\sigma}_\epsilon(CES )$ &0.7643&0.7815&0.7895  \\ \hline
minima& $\hat{\sigma}_{LSratio}$ &0.1552 &0.1533 &0.1568 \\
(CES Method)&$\hat{\sigma}_{ratio}$  &0.1548 &0.1533 &0.1567 \\ \hline
2a Method& $\hat{\sigma}_{\epsilon}(LS2)$ &0.8009 &0.7991 &0.8026 \\
&$\hat{\sigma}_{\epsilon}(MS)$  &0.7990 &0.7988 &0.8024 \\ \hline
standard & $\hat{\sigma}_x(LS )$ & 4.8885 & 5.0252 & 4.9894 \\
deviations & $\hat{\sigma}_x(CES)$ & 5.3632 & 5.2682 & 5.1372 \\ \hline

\hline
\multicolumn{5} {|c|} {  $\rho=0.0000$  $\beta=0.0000$ $\sigma_{\epsilon}=1.9000$    } \\
\multicolumn{5} {|c|} {$\sigma_x=1.5000$ $\sigma_y=1.9000$ }\\  \hline
  & nsim=100 & n=20& n=50 & n=100 \\ \hline
slopes &$\hat{\beta}(LS)$& -0.0341& -0.0026& -0.0073 \\
&$\hat{\beta}(MS)$& -0.0498& 0.0016 & -0.0093 \\ \hline
minima& $\hat{\sigma}_\epsilon(LS )$ &1.8342&1.8437&1.8676\\
(LS Method)& $\hat{\sigma}_\epsilon(CES )$ &1.8385&1.8445& 1.8678  \\ \hline
minima& $\hat{\sigma}_{LSratio}$ &1.2130 &1.2152 &1.2327 \\
(CES Method)&$\hat{\sigma}_{ratio}$  &1.2095 &1.2147 &1.2326 \\ \hline
2a Method& $\hat{\sigma}_{\epsilon}(LS2)$ &1.9083 &1.8880 &1.8974 \\
&$\hat{\sigma}_{\epsilon}(MS)$  &1.9029 &1.8872&1.8972 \\ \hline
standard & $\hat{\sigma}_x(LS )$ & 1.4768 & 1.4924 & 1.5036 \\
deviations & $\hat{\sigma}_x(CES)$ & 1.6151 & 1.5656 & 1.5473 \\ \hline

\hline
\multicolumn{5} {|c|} {  $\rho=0.5727$  $\beta=0.8008$ $\sigma_{\epsilon}=1.7659$    } \\
\multicolumn{5} {|c|} {$\sigma_x=1.5403$ $\sigma_y=2.1541$ }\\  \hline
  & nsim=100 & n=20& n=50 & n=100 \\ \hline
slopes &$\hat{\beta}(LS)$& 0.8156& 0.7890& 0.7881 \\
&$\hat{\beta}(MS)$& 0.8197& 0.7853 & 0.7911 \\ \hline
minima& $\hat{\sigma}_\epsilon(LS )$ &1.6520 &1.7236  & 1.7692\\
(LS Method)& $\hat{\sigma}_\epsilon(CES )$ &1.6566 & 1.7251& 1.7695  \\ \hline
minima& $\hat{\sigma}_{LSratio}$ &1.0296 &1.1276 &1.1376 \\
(CES Method)&$\hat{\sigma}_{ratio}$  &1.0266 &1.1268 &1.1375 \\ \hline
2a Method& $\hat{\sigma}_{\epsilon}(LS2)$ &1.7315 &1.7614 &1.7974 \\
&$\hat{\sigma}_{\epsilon}(MS)$  &1.7266 &1.7602 & 1.7972 \\ \hline
standard & $\hat{\sigma}_x(LS )$ & 1.5720 & 1.5078 & 1.5417 \\
deviations & $\hat{\sigma}_x(CES)$ & 1.7254 & 1.5796 & 1.5859 \\ \hline

\end{tabular}
\end{center}
\end{table}

\section{R Code }
The R code for the functions needed to let any reader easily reproduce the analysis and extend the ideas to other correlation coefficients is presented here. The R function \textit{Pfcn} specifies how the variable $b$ is to be estimated using Pearson's CC. General use by other CCs is done by defining \textit{rfcn} to be the CC used in \textit{rtest} which sets up regression equation (1) and its solution using \textit{uniroot}.  So the CC choice is done by setting \textit{rfcn} to be \textit{Pfcn} when Pearson is desired. Then  \textit{rtest} gives the objective function for  \textit{optimize}, called for by \textit{outces}, which defines the data and does the iterations to minimize $s$ in \textit{rtest}.

 $\quad Pfcn = function(b,x,y)  \{cor(x,y-b*x)  \}$

 $ \quad rfcn =  Pfcn$

$\quad rtest = function(b,x,y) \quad \{y1 = sort(y - b*x)$

$\quad s = uniroot(rfcn,c(-4,4),x=xsr, y=y1)\$ root, return(s)\} $
Quantity  $xsr = sort(x)$, the sorted $x$ values, is used in \textit{uniroot} within  \textit{rtest}.

$\quad outces = optimize(rtest, c(-5,5), x=x, y=y)$

\noindent $outces\$min$ is the slope estimate, $\hat{\beta}$, for the regression  and $outces\$obj$ is the CES minimum, the MS result, $\hat{\sigma}_{ratio}$.
 If $cres$ equals the vector of MS residuals, $\hat{\sigma}_{\epsilon}(CES)=sqrt(var(cres))$.

$\hat{\sigma}_{\epsilon}(LS)$ is the estimate of $\sigma_{\epsilon}$ using the linear model R routine, $lm$. If $lres$ equals the vector of LS residuals, $\hat{\sigma}_{\epsilon}(LS)=sqrt(var(lres))$.

Let $ysr$ be the sorted values of the LS regression residuals, from $lm$ function. Then

$\quad \hat{\sigma}_{LSratio} = uniroot(rfcn,c(0,12),x=xsr,y=ysr)$.

$\quad p3 = (1:n)/(n+1)$; $q3 = qnorm(p3)$

$\quad \hat{\sigma}_x(CES) = uniroot(rfcn,c(0,15),x=q3,y=xsr)\$root $.
This gives the CES estimate of $\sigma_x$.

$\quad \hat{\sigma}_{\epsilon}(LS2)=\hat{\sigma}_{LSratio}*\hat{\sigma}_x(CES)$ and $\hat{\sigma}_{\epsilon}(MS)=\hat{\sigma}_{ratio}*\hat{\sigma}_x(CES)$.
$\quad \hat{\sigma}_x(LS) $ is $sqrt(var(x))$

Some possible CCs are listed previously. As an example, $GDfcn$ is defined like $Pfcn$ but with  $cor$ replaced by $GDave$, the R-routine for GDCC as found in \textit{Cordef\&reg.R}.

.

\begin{table}
	\caption*{Table 2: Comparison of Seven Minimization Processes}
	\vspace*{-.3cm}
	\centerline{All from One Sample}
	\vspace*{.5cm}

		\begin{tabular}{|l|c|c|c|}
			\hline
		\multicolumn{4} {|c|} {  $\rho=0.5727$  $\beta=0.8008$ $\sigma_{\epsilon}=1.7659$    } \\
		\multicolumn{4} {|c|} {n = 25 $\sigma_x=1.5403$ $\sigma_y=2.1541$ }\\  \hline
		&  $\hat{\beta}(LS)$ & $\hat{\sigma}_\epsilon(LS )$ & $\hat{\sigma}_{\epsilon}(LS2)$ \\
		& $\hat{\beta}(CES)$ &	  $\hat{\sigma}_\epsilon(CES )$ & $\hat{\sigma}_{\epsilon}(MS)$ \\  \hline
			LS  & 1.0086& 1.6799& 1.7592 \\
			Pearson& 1.0333& 1.6804& 1.7591 \\ \hline
	       LS2&    &     & 1.9529\\
			GDCC&  0.8228 &1.7058& 1.6699 \\ \hline
			 LS2&       &     & 1.8013 \\
			Kendall&1.1598 &  1.6971& 1.7441 \\\hline
			 LS2&       &     & 1.8546 \\
           Gini &  1.2021&1.7080 & 1.7916 \\\hline
            LS2&       &     & 1.7565 \\
           Absolute &0.8617 &  1.6961& 1.7222\\ \hline
           LS2&       &     & 1.8975\\
           MAD &  0.6862 & 1.7568 & 1.7255   \\ \hline
			
		\end{tabular}
	
\end{table}

\section{Conclusion}
In LS minimization and zero correlation of $x$ with the residuals imply each other. This is not true for CES. The zero method is shown in Gideon (2012) or in Gideon and Rummel (1992).
To change equation (1) to include multiple linear regression, add additional linear terms.
For example, a second regression variable is added using $b_2\underline{x}_2$. Now, however, the term $(\underline{y}-b_1\underline{x}_1-b_2\underline{x}_2)^0$  needs to be regressed against $\underline{y}^0$, which can be accomplished  by varying $b_1$ and $b_2$ to minimize $s$. This $s$ estimates $\sigma_\epsilon/\sigma_y$. Thus CES  maximizes  $1-{\sigma}_{\epsilon}^{2}/{\sigma}_{y}^{2}$, the multiple correlation coefficient.   Gideon (2012) contains this extension and others, focusing on the Absolute Value and Greatest Deviation correlation coefficients. The author's conjecture is that using Pearson's $r_p$ to find the CES minimum is equivalent to the usual least squares method. This conjecture has not been studied theoretically. Is there a proof?

There are three main reasons for this paper: first, to show that the Minimum Slope criterion of CES  using Pearson's correlation coefficient is apparently as good as the least squares criterion in simple linear regression. Second, to show the R commands that allow the use of any CC in  place of $r_p$ so as to offer a very general estimation system, the Correlation Estimation System. Third, because of the first two, the question for model building becomes not just which is the "best" model but also which is the "best" criterion to select the model. All the models fit by CES with other correlation coefficients in Gideon (2012) were outstanding. Some CES  distribution theory was given in Gideon (2010). The generality of CES makes it easy to implement in a wide variety of regression situations. One only needs the R program \textit{Cordef\&reg.R} (or existing R routines for Pearson, Spearman, and Kendall) to set up the correlation coefficients and the regression sequence. One may find that the classical fit is not the "best".

\section{References}

\setlength{\parindent}{0in}
\setlength{\parskip}{.05in}
Author website: hs.umt.edu/math/people/default.php?s=Gideon

Gideon, R.A. (2007). The Correlation Coefficients, \textit{Journal of Modern Applied
Statistical Methods}, \textbf{6}, no. 2, 517--529.

Gideon, R.A. (2010). The Relationship between a Correlation Coefficient and its Associated Slope
Estimates in Multiple Linear Regression, \textit{Sankhya}, \textbf{72}, Series B, Part 1, 96--106.

Gideon, R.A. (2012). Obtaining Estimators from Correlation Coefficients: The Correlation Estimation System and R, \textit{Journal of Data Science}, \textbf{10}, no. 4, 597--617.

Gideon, R.A. and Hollister, R.A. (1987). A Rank Correlation Coefficient Resistant to
Outliers, \textit{Journal of the American Statistical Association}, \textbf{82}, no. 398, 656--666.

Gideon, R.A., Prentice, M.J., and Pyke, R. (1989). The Limiting Distribution of the Rank
Correlation Coefficient $r_{gd}$. In: Contributions to Probability and Statistics (Essays in
Honor of Ingram Olkin), ed. Gleser, L.,J., Perlman, M.D., Press, S.J., and Sampson, A.R.
Springer-Verlang, N,Y., 217--226.

Gideon, R.A. and Rothan, A.M., CSJ (2011). Location and Scale Estimation with Correlation
Coefficients, \textit{Communications in Statistics-Theory and Methods}, \textbf{40}, Issue 9, 1561--1572.

Gideon, R.A. and Rummel, S.E. (1992). Correlation in Simple Linear Regression, unpublished
paper (http://www.math.umt.edu/gideon/ \\ CORR-N-SPACE-REG.pdf), University of Montana, Dept.
of Mathematical Sciences.

Gini, C. (1914). L'Ammontare c la Composizione della Ricchezza della Nazioni,\textit{ Bocca},
Torino.

Randles, R. H. and Wolfe, D. A. (1979). Introduction to the Theory of Nonparametric Statistics,
Wiley \& Sons, New York.

Rummel, Steven E. (1991). A Procedure for Obtaining a Robust Regression Employing the Greatest
Deviation Correlation Coefficient, Unpublished Ph.D. Dissertation, University of Montana,
Missoula, MT 59812, full text accessible through UMI ProQuest Digital Dissertations.

Sheng, HuaiQing (2002). Estimation in Generalized Linear Models and Time Series Models with
Nonparametric Correlation Coefficients, Unpublished Ph.D. Dissertation, University of Montana,
Missoula, MT 59812, full text accessible through\\
http://wwwlib.umi.com/dissertations/fullcit/3041406.

\end{document}